\documentclass[superscriptaddress,showpacs,floatfix,prl,twocolumn]{revtex4}
\usepackage{amssymb}

\usepackage{graphicx}
\usepackage{amsmath}

\newcommand{\be}{\begin{equation}}
\newcommand{\ee}{\end{equation}}

\newcommand{\kk}{{\mathbf k}}
\newcommand{\eq}[1]{(\ref{#1})}

\newcommand{\mume}{\mu \rm m}
\newcommand{\im}{\textrm{Im}}
\newcommand{\re}{\textrm{Re}}

\newcommand{\rr}{{\bf r}}

\begin{document}

\title{Spatial and spectral shape of\\ inhomogeneous non-equilibrium exciton-polariton condensates} 
\author{Michiel Wouters}
\affiliation{TFVS, Universiteit Antwerpen, Groenenborgerlaan 171, 2020
  Antwerpen, Belgium} 
\author{Iacopo Carusotto}
\affiliation{BEC-CNR-INFM and Dipartimento di Fisica, Universit\`a di
  Trento, I-38050 Povo, Italy}
\author{Cristiano Ciuti}
\affiliation{Laboratoire Mat\'eriaux et Ph\'enom\`enes Quantiques, Universit\'e Paris Diderot, Paris VII, 75205 Paris Cedex 13, France}
\begin{abstract}
We develop a mean-field theory of the spatial profile and the spectral properties of polariton condensates in nonresonantly pumped semiconductor microcavities in the strong coupling regime. Predictions are obtained for both the continuous-wave and the pulsed excitation regimes and the specific signatures of the non-equilibrium character of the condensation process are pointed out. A striking  sensitivity of the condensate shape on the optical pump spot size is demonstrated by analytical and numerical calculations, in good quantitative agreement with recent experimental observations.
\end{abstract}
\pacs{
03.75.Kk,  	 % Bose-Einstein condensation dynamic properties 
71.36.+c, 	 % Polaritons 
42.65.Sf, 	 % Dynamics of nonlinear optical systems; optical
             % instabilities, optical chaos and complexity, and
             % optical spatio-temporal dynamics
05.70.Ln		 % Nonequilibrium and irreversible thermodynamics
}
\maketitle

%\emph{Introduction}-- 
First evidences of Bose-Einstein condensation (BEC) in a solid-state system have been recently reported in a gas of exciton-polaritons in a semiconductor microcavity in the strong coupling regime \cite{maxime,maxime-large,kasprzak,yamamoto,snoke,christopoulos}.
In addition of being a remarkable example of an exciton condensate, this system opens interesting perspectives towards the study of the BEC phenomenon in completely new regimes.
Polariton condensates differ in several fundamental aspects from the ideal case generally considered in textbooks: polariton-polariton interactions are significant and the
system is far from thermodynamical equilibrium.
While the separate effect on condensation of the interactions and of the non-equilibrium condition is already well understood from either ultracold atom \cite{bec-book} or laser \cite{laser} theory, not much is yet known about the interplay of the two effects when simultaneously present.
In this case, the Bose gas is in fact a quantum degenerate, interacting many-body system whose stationary state does not correspond to a thermal equilibrium state, but rather originates from a dynamical balance of pumping and losses~\cite{nostri,goldstone,nonresonant}

Some striking consequences of the non-equilibrium condition on the elementary excitations have been recently predicted, the propagating sound mode of equilibrium condensate being e.g. replaced by a diffusive mode~\cite{goldstone,nonresonant}.
Even more remarkably, an unexpectedly rich behavior has been observed in recent experiments in the spatial and spectral shapes of the condensate depending on the size of the pump spot~\cite{maxime,maxime-large}.
While at equilibrium, BEC generally occurs at zero momentum and the trapping potential is only responsible for the $k$-space broadening due to finite size~\cite{bec-book}, the first experimental studies of polariton condensates performed with a relatively small pump laser spot showed Bose-Einstein condensation into a ring of momentum states with a non-zero wave vector~\cite{maxime}; mutual coherence of different $k$-states was however interferometrically demonstrated, which proved that a true condensate was created, and not a fragmented one~\cite{leggett}. Standard condensation around $k=0$ was then recovered in later experiments using a much wider excitation spot~\cite{maxime-large,kasprzak}.
So far, most of these experimental observations have challenged theoretical understanding~\cite{comment_1}: the purpose of the present Letter is to propose a complete and unified theoretical model able to explain them in a simple and physically transparent way.

Our work is based on a mean-field study of non-equilibrium condensates which takes explicitly into account the effect of the spatially finite pump spot by means of a generalized Gross-Pitaevskii equation (GPE) as developed in~\cite{nonresonant}. 
This approach can be used to theoretically model non-equilibrium Bose-Einstein condensation for any pump laser geometry and for any microcavity disorder potential.
As long as mean-field is valid, solving the GPE provides complete information on the shape of the polariton condensate in both real- and momentum-space, as well as on its spectral properties under a pulsed excitation. The formal similarities with the equations appearing in the context of pattern formation in nonlinear dynamical systems far from equilibrium~\cite{cross}, in particular hydrodynamical~\cite{chomaz} and nonlinear optical~\cite{kuszelewicz} ones will be of great utility to obtain a physical understanding of the complicate spatial structures that appear as a consequence of the interplay of inhomogeneity, nonlinearity, driving and dissipation. 

At mean-field level, the dynamics of the condensate macroscopic wavefunction $\psi(\rr)$ is described by a generalized Gross-Pitaevskii equation~\cite{nonresonant} of the form:
\begin{multline}
i\hbar \frac{\partial \psi(\rr)}{\partial t}=\Big\{E_0 -\frac{\hbar^2}{2m} \nabla_\rr^{2}+
\frac{i\hbar}{2}\big[R[n_R(\rr)]-\gamma_c\big] \\
+V_{ext}(\rr)+\hbar g\,|\psi(\rr)|^{2}+V_R(\rr) \Big\}
\psi(\rr), \label{eq:GP}
\end{multline}
where $E_0=\hbar\omega_0$ and $m$ are respectively the minimum and the effective mass of the lower polariton branch and  $g>0$ quantifies the strength of repulsive binary interactions between condensate polaritons. Whenever needed, cavity disorder can be included as an external potential term $V_{ext}(\rr)$.
Condensate polaritons have a linear loss rate $\gamma_c$, and are continuously replenished by stimulated emission from the polariton reservoir created by the nonresonant optical pump. At the simplest level, the corresponding gain rate $R[n_R]$ can be described by a monotonically growing function of the local density $n_R(\rr)$ of reservoir polaritons in the so-called bottleneck region~\cite{porras}. At the same time, the reservoir produces a mean-field repulsive potential $V_R(\rr)$ that can be approximated by the linear expression $V_R(\rr)\simeq \hbar g_R\,n_R(\rr)+\hbar {\mathcal G}\,P(\rr)$, where $P(\rr)$ is the (spatially dependent) pumping rate and $g_R,{\mathcal G}>0$  are phenomenological coefficients to be extracted from the experiment~\cite{footnote2}.
The GPE equation \eq{eq:GP} for the condensate has then to be coupled to a rate equation for $n_R(\rr)$
\begin{equation}
\dot{n}_R(\rr)=P(\rr)-\gamma_R\,n_R(\rr)-R[n_{R}(\rr)]\,|\psi(\rr)|^{2}:\label{eq:rate}
\end{equation}
polaritons are injected at a rate $P(\rr)$ and relax at an effective rate $\gamma_R\gg \gamma_c$~\cite{footnote1}. Depletion of the reservoir density due to the stimulated emission into the condensate mode is taken into account by the $R[n_R(\rr)]\,|\psi(\rr)|^2$ term. 

%{\em Stationary state}--
In the homogeneous case, i.e. under a uniform pumping and in the absence of any external potential, the equations (\ref{eq:GP}-\ref{eq:rate}) admit simple analytical stationary solutions. 
Below the threshold, the condensate density remains zero $|\psi|^2=0$, while the reservoir one  grows linearly with the pump intensity, $n_R= P/\gamma_R$.
At the threshold pump intensity $P^{th}$, the stimulated emission rate exactly compensates the losses $R[n_R^{th}]=\gamma_c$ and the empty condensate solution with $\psi=0$ becomes dynamically unstable.
Above the threshold, the reservoir density is homogeneous $n_{R}(\rr)=n_{R}$ and the condensate wavefunction is of the form $\psi(\rr)=e^{i(\kk_c\cdot \rr-\omega_c t)}\,\psi_0$.
Inserting this ansatz into the motion equation, we find that $n_R$ is clamped at the threshold value $n_R^{th}$, while the condensate density grows as $|\psi_0|^2=(P-P^{th})/\gamma_c$. The condensate wave vector $\kk_c$ remains so far undetermined and stable solutions with arbitrary values of $\kk_c$ can be found. Finally, the oscillation frequency $\omega_c$ is fixed by the state equation:
\be
\omega_c-\omega_0 =\frac{\hbar\,k_c^2}{2m}+g\,|\psi_0|^2+g_R\,n_R+\mathcal{G}\,P.
\label{eq:mustat}
\ee

The physics is much richer in the presence of an inhomogeneous intensity profile $P(\rr)$ of the pump. 
In this case, we can still look for stationary solutions of the form:
\begin{eqnarray}
\psi(\rr,t)&=&\psi_0(\rr)\,e^{-i\omega_c t}=\sqrt{\rho(\rr)}\,e^{i(\phi(\rr)-\omega_c t)} \label{eq:psi_stat} \\ 
n_{R}(\rr,t)&=& n_{R}(\rr). \label{eq:nr_stat}
\end{eqnarray}
Here, $\rho(\rr)$ and $\phi(\rr)$ are the local density and phase of the condensate wavefunction $\psi(\rr)=\sqrt{\rho(\rr)}\,\exp[i\phi(\rr)]$.  The condensate frequency $\omega_c$ is the same at all points, while the local condensate wave vector $\kk_c(\rr)$ is defined as the spatial gradient of the condensate phase $\kk_c(\rr)=\nabla_\rr \phi(\rr)$.
Stationarity of the solution then imposes that 
\begin{eqnarray}
\hbar\omega_c &=& \omega_0+\frac{\hbar^2 k_c^{2}}{2m}+V_{ext}+\frac{\hbar^2}{2m} \frac{\nabla_\rr^2\sqrt{\rho}}{\sqrt{\rho}}
\nonumber \\
&+&\hbar g\rho+\hbar g_R n_R+\hbar {\mathcal G} P  \label{eq:mu_eig}  \\
0 &=& \big[R[n_R]-\gamma_c\big]\rho -\frac{\hbar}{m}\,\nabla_\rr \cdot( \rho \kk_c) \label{eq:mu_imag} \\
P&=&\gamma_R\,n_R +R[n_R]\,\rho .
\end{eqnarray}

Provided the spatial variation of the pump profile $P(\rr)$ is smooth enough, one can perform a kind of Local Density Approximation (LDA), where the quantum pressure term (proportional to $\nabla_\rr^2 \sqrt{\rho}$) in \eq{eq:mu_eig} and the current divergence term in \eq{eq:mu_imag} are neglected.
Within this LDA approach, the condensate density  vanishes $\rho(\rr)=0$ at all points $\rr$ where the pump intensity is below the threshold $P(\rr)<P_{th}$. As the oscillation frequency $\omega_c$ of the condensate is spatially constant, the variation of $P$ across the pump spot must be compensated by a corresponding spatial variation of the local wave vector $\kk_c$.

For a circular pump profile $P(r)$ and in the absence of disorder $V_{ext}=0$, the condensate frequency $\omega_c$ is determined by the condition that the local condensate wave vector vanishes $\kk_c=0$ at the center of the spot: $\omega_c-\omega_0=g\,\rho(r=0) +g_R\, n_R(r=0)+\mathcal{G}\,P(r=0)$.
For standard (e.g. gaussian) pump spots with a monotonically decreasing intensity profile along the radial direction, $\kk_c$ is in the outward radial direction and its modulus $k_c$ monotonically grows in the radial direction, and reaches its maximal value at the condensate edge where $P(r)=P_{th}$.
The repulsive interactions create in fact an antitrapping potential $V_{at}(r)=\hbar g\,\rho(r)+\hbar g_R\, n_R(r)+\hbar \mathcal{G}\,P(r)$ which ballistically accelerates the condensate polaritons away from the center and the state equation \eq{eq:mustat} simply describes the conservation of the mechanical energy during the flow.
In order to verify the robustness of these analytical considerations and assess their validity for the typical values of the experimental parameters, extensive numerical simulations of the GPE equation \eq{eq:GP} coupled to the reservoir evolution equations (\ref{eq:rate}) have been performed for a wide range of pump parameters.

\begin{figure}[htbp]
\begin{center}
\includegraphics[width=1.\columnwidth,angle=0,clip]{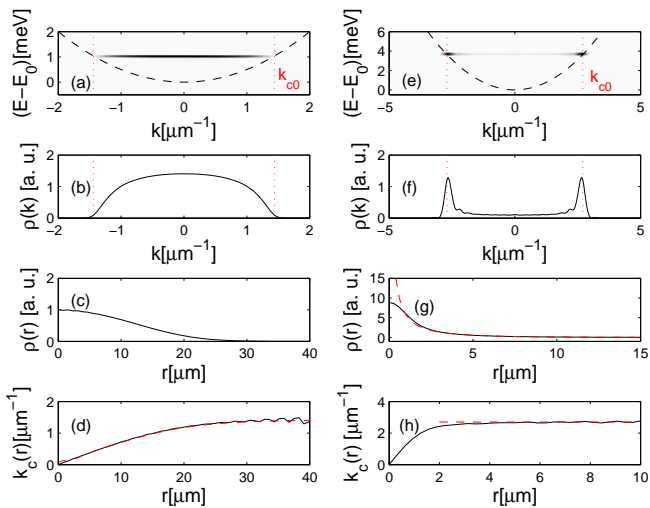} %nonres/maxime/anal/
\end{center}
\caption{(Color online) Numerical results in the absence of disorder for a circular excitation pump spot. The polariton distribution is shown for a large $\sigma_p=20\,\mume$ (a-d) and a small $\sigma_p=1\,\mume$ (e-h) pump laser spot. Panels (a,e) give the $(k,E)$ emission pattern, (b,f) the polariton distribution in reciprocal space, (c,g) in real space and (d,h) the local wave vector $k_c(r)$. The wave vector $k_{c0}$ of a free polariton at  $\omega_c$ is indicated by the dotted lines in (a,b,e,f); the dashed line in (g) is the analytical approximations to the density tail, and the dashed line in (d,h) are the LDA predictions to the local wave vector. 
All quantities in real (momentum) space depend only on the radial coordinate $r=|\rr|$ ($k=|\kk|$). 
Values of parameters used in the simulations: $\hbar g=0.03\, \textrm{meV} \,\mume^2$, $\hbar\gamma_c=0.5\,\textrm{meV}$, $\hbar\gamma_R=2\,\textrm{meV}$, $\hbar R[n_R]=(0.1 \textrm{meV} \mume^2) \times n_R$  $\hbar\,g_R=0$, $\mathcal G = 0.035 \mume^2$ and $P/P_{th}=2$ (left panels), $\mathcal G=0$ and $P/P_{th}=48$ (right panels).}
\label{fig:ideal}
\end{figure}

The case of a cw pump with a wide Gaussian spot of waist $\sigma_p=20\,\mu m$ is shown in the four plots in the left column of Fig.~\ref{fig:ideal}. 
The maximum height of the antitrapping potential $V_{at}$ is chosen in a way to reproduce the experimentally observed  blue-shift of the emission frequency $\omega_c-\omega_0$ of the order of $1\,\textrm{meV}$~\cite{maxime-large}.
As expected, the long-time dynamics of the system tends to a steady state with a single oscillation frequency $\omega_c$ and stationary reservoir densities.
In agreement with the previous analytical discussion, the $k$-space distribution [Fig.~\ref{fig:ideal}(b)] is contained in the $k<k_{c0}$ region ($\hbar k_{c0}^2/2m =\omega_c-\omega_0$) delimited by the free particle dispersion (vertical dashed lines in the figure). Still, the broadening of the condensate momentum distribution due to the ballistic acceleration largely exceeds what predicted by mere finite size effects. This in stark contrast with the case of equilibrium condensates, where the condensate phase is uniform throughout the whole cloud and the local wave vector is thus everywhere $\kk_c=0$~\cite{bec-book}. To confirm our interpretation, we have compared the numerical result for the local wave vector with the LDA prediction. The two are plotted in Fig.~\ref{fig:ideal}(d) as respectively a full and a dashed line: the agreement between the two is everywhere excellent.

\begin{figure}[htbp]
\begin{center}
\includegraphics[width=\columnwidth,angle=0,clip]{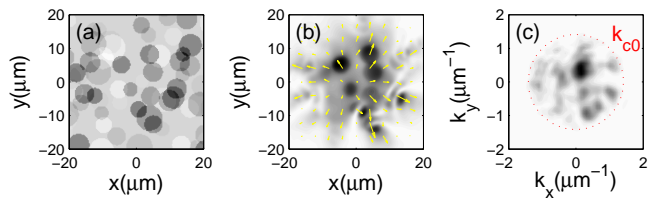} %nonres/maxime/anal/
\end{center}
\caption{(Color online) Numerical results with disorder. (a) Disorder potential (gray scale with 1meV range) , polariton distribution in 2D (b) real and (c) momentum space . The arrows in (b) indicate the polariton current. The radius of the dotted circle in (c) is the free polariton wave vector $k_{c0}$.}
\label{fig:disorder}
\end{figure}

Although this simple calculation is able to correctly account for the main experimental observations, some features are still missing. As one can see in Fig.8.9 of Ref.\cite{maxime-thesis}, some structure is in fact present on top of the broad profile shown in Fig.~\ref{fig:ideal}(a,b), in particular a narrow peak centered at a finite $\kk\neq 0$ momentum. 
To explain these features we have performed GPE simulations including a weak disorder potential $V_{ext}(\rr)$ acting on polaritons. Fig.\ref{fig:disorder} gives an example of the outcome for a given realization of the disorder potential, constructed as an ensemble of circular defects [Fig.\ref{fig:disorder}(a)]. The $\kk$-space emission shown in Fig.\ref{fig:disorder}(c) is still contained in the $k<k_{c0}$ circle as in the clean system case, but a significant speckle-like modulation appears on top of the regular background, with a few dominating narrow peaks. 
This phenomenology can be understood in terms of the interference of the emission from the different spatial positions: in agreement with experimental observations, the spatial profile of the condensate [Fig.\ref{fig:disorder}(b)] consists in fact of several, yet mutually coherent spots connected by lower density regions. In particular, the disorder is not able to spoil the outward radial flow pattern which, despite some local distortions, remains clearly visible in the figure and is responsible for the $\kk\neq 0$ position of the peaks in the $\kk$-space pattern.

The properties of the polariton condensate are completely different if a small excitation spot is used.
Even though the local density approximation can no longer be performed, exact analytical predictions for the condensate wave function can still be obtained in the region far outside the pump spot. In this region, the condensate density is very small and polaritons no longer feel the repulsive potential, so that the radial part of the GPE \eq{eq:GP} reduces to a linear Helmholtz equation of the form $(\hbar/{2m})\,\nabla_\rr^2 \psi +(\omega_c-\omega_0+i\gamma_c/2)\psi=0$,
whose converging solution for $r\rightarrow \infty$ has to be chosen. The frequency $\omega_c$ is fixed by the dynamics in the excitation region, and its analytical determination goes beyond the scope of the present work.
The solution to the Helmholtz equation has the simple analytic expression
$\psi(r)=H^{(1)}_0(\tilde{k}_c r)$,
in terms of the first kind Hankel function $H^{(1)}_\nu$. The complex wave vector is $\tilde{k}_c=\sqrt{2m(\omega_c-\omega_0+i\gamma_c/2)/\hbar}$. 
From the asymptotic expansion of $H^{(1)}_\nu$, one immediately gets to the exponential decay $\psi(r\rightarrow \infty)\propto \exp(-\kappa_{pen}r)/\sqrt{r}$ with a spatial rate $\kappa_{pen}=\im[\tilde{k}_c]\simeq \gamma_c/(2v_c)$, and a propagation speed $v_c=\hbar k_c/m$, with $k_c=\re[\tilde{k_c}]=\sqrt{2m(\omega_c-\omega_0)/\hbar}$~\cite{footnote3}.
These analytical considerations are accurately confirmed by a full numerical integration of (\ref{eq:GP}-\ref{eq:rate}) for a small pump spot $\sigma_p=1\,\mu\textrm{m}$, whose results are summarized in the plots in the right column of Fig. \ref{fig:ideal}.
As the pump laser spot $\sigma_p$ is much smaller than the characteristic propagation length $\kappa_{pen}^{-1}$, most of the polaritons are ballistically propagating outside the excitation region. The $k$-space emission then concentrates on the ring of radius $k_c$; the contribution from the excitation region provides the weak pedestal at $k<k_c$, while almost no emission is present for $k>k_c$. The $(k,E)$ pattern is analogously peaked around $(\pm k_{c0},\hbar\omega_c)$ and shows a weaker pedestal on the horizontal segment at $E=\hbar\omega_c$ contained inside the free polariton dispersion.
Qualitative agreement of these results with experimental observations in Fig.1 of Ref.~\cite{maxime} is excellent: the ring structure is recovered as well as the relation between the emission frequency $\omega_c$ and the wave vector $k_c$. 

\begin{figure}[tbp]
\begin{center}
\includegraphics[width=0.5\columnwidth,angle=0,clip]{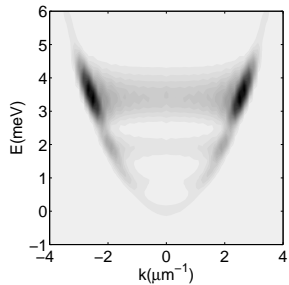} 
%nonres/maxime/anal/
\end{center}
\caption{$(k,E)$ emission pattern from a simulation with a small pump spot ($\sigma_p=1\,\mume$) and the following time profile: linear rise time of $1.3\,\textrm{ps}$, plateau and fall time of $2.6\,\textrm{ps}$.}
\label{fig:pulse}
\end{figure}

As a final point of the Letter, it can be interesting to investigate the effect of a finite duration of the excitation pulse.
The result of a numerical simulation of the GPE model~\cite{footnote4} is shown in Fig. \ref{fig:pulse}. Although the condensate does not have time to reach its stationary state, its basic properties remain similar to the ones shown in the right column of Fig.\ref{fig:ideal}. The main difference is the appearance  in the $(k,E)$ emission pattern of additional horizontal lines corresponding to new emission frequencies.
It is interesting to note that this feature is in remarkable qualitative agreement
with the experimental observations shown in fig.2b of Ref.~\cite{maxime}.

In conclusion, we have developed a full theory of the spatial and spectral shape of polariton Bose-Einstein condensates which is able to explain all the peculiar features observed in recent experiments~\cite{maxime,maxime-large,maxime-thesis}.
Significant differences are pointed out with respect to the standard case of condensates at equilibrium: because of the repulsive interactions, the steady-state is characterized by a significant flow in the outward radial direction and polaritons can ballistically propagate outside the excitation spot.
The agreement of the results of our generalized Gross-Pitaevskii approach with experiments is promising for the application of the present model to more complex geometries that may be relevant in the design of polariton lasers based on suitably patterned structures. From the theoretical point of view, the next step will be inclusion of fluctuations around the mean-field, so as to obtain quantitative predictions for the condensate coherence properties and, eventually, for the critical properties in the vicinity of the non-equilibrium phase transition.

Stimulating discussions with M. Richard, A. Baas, D. Le Si Dang, D. Sarchi, V. Savona, M. Szyma\'nska, J. Tempere and J. Devreese are greatfully acknowledged. This research has been supported financially by the FWO-V projects Nos.G.0356.06, G.0115.06 and the Special Research Fund of the University of Antwerp, BOF NOI UA 2004.M.W. acknowledges financial support from the FWO-Vlaanderen in the form of a ``mandaat  Postdoctoraal Onderzoeker''. 
IC acknowledges hospitality at the Centre Emile Borel of the Institut Henri Poincar\'e and financial support from CNRS. 

Shortly before submission of this manuscript, the preprint~\cite{keeling} has appeared where similar issues are addressed from a slightly different perspective.


\begin{thebibliography}{99}
\bibitem{footnote2} Although the present knowledge of these parameters is quite poor, we have checked that the conclusions of our analysis do not depend on the specific values actually used in the simulations. The only quantity that matters is the total blue-shift $\omega_c-\omega_0$ of the condensate frequency, a quantity which is easily extracted from experimental data.
\bibitem{footnote1} Although the loss rate of bottleneck polaritons is generally of the same order of magnitude as the condensate one, the effective relaxation rate $\gamma_R$ is much faster thanks to thermalization of bottleneck polaritons with the much larger bath of high energy excitations created by the pump laser. A detailed discussion on these issues is postponed to a forthcoming publication.
\bibitem{footnote3} Note that the limit $\gamma_c\rightarrow 0$ of the present theory does not correspond at all to the equilibrium state of trapped condensates. In a broad sense, it rather reminds the physics of the expansion after the trap potential has been switched off~\cite{bec-book}.
\bibitem{footnote4} In the simulations, a pulse duration much longer than the real laser duration has been considered so to account in an effective way for the complex relaxation dynamics of the excited carriers.
\end{thebibliography}
\end{document}